\documentstyle[12pt]{article}
\def\be{\begin{equation}}
\def\ee{\end{equation}}
\def\bea{\begin{eqnarray}}
\def\eea{\end{eqnarray}}

\begin{document}

\begin{flushright}
IPM/P-2002/014\\
\end{flushright}

\begin{center}
{\Large{\bf Space-Time Symmetries, T-Duality and Gauge Theory }}                  
										 
\vskip .5cm   
{\large Davoud  Kamani}
\vskip .1cm
 {\it Institute for Studies in Theoretical Physics and
Mathematics (IPM)
\\  P.O.Box: 19395-5531, Tehran, Iran}\\
{\sl E-mail: kamani@theory.ipm.ac.ir}
\\
\end{center}

\begin{abstract} 

In this paper we study $U(1)$ gauge 
transformations on the space-time coordinates and on the background
fields $g_{\mu\nu}$ and $\phi$.
For some special gauge functions, gauged coordinates and gauged $U(1)$
field are equivalent to the rotated coordinates and rotated gauge field.
We find gauge transformations that are symmetries of the string action.
Also we obtain general $T$-duality transformations for the background
fields. For special background fields this duality is equivalent to a gauge 
transformation.

\end{abstract} 
\vskip .5cm


\newpage
\section{Introduction}
 
As it has been mentioned by A.M. Polyakov, 
the relation between gauge fields and 
strings is very important \cite{1}. In fact this relation 
is useful in solving the problem of gauge-string-spacetime correspondence. 
From the other side $T$-duality transformation is an exact symmetry
of the closed string theory \cite{2}. We shall concentrate on the gauge
theory, string theory and $T$-duality. 
In the presence of a $U(1)$ gauge field,
we study the gauge transformations of the spacetime coordinates,
the metric $g_{\mu\nu}$ and the dilaton $\phi$. We 
compare these with some other transformations of the spacetime. In other
words some of the transformations on the spacetime and gauge fields can 
be interpreted as gauge transformations. We find gauge transformations that
are symmetries of the superstring action.

Previously we have studied a general $T$-duality \cite{3}. By this,
we find general $T$-duality transformations on the spacetime coordinates
and on the NS$\otimes$NS fields. For some special background fields this
general $T$-duality is equivalent to a gauge transformation.
In this case the above dual space is a gauged space. In this gauged space 
we find the effective field strength that can be seen by a closed string.

In the second section, we shall study gauge transformations of the
spacetime and their effects on the string action. In the third section, 
using the above general $T$-duality, we concentrate on the 
gauge-string relations.
%%%%%%%%%%%%%%%%%%%%%%%%%%%%%%%%%%%%%%%%%%%%%%%%%%%%%%%%%%%%%%%%%%%%%%%%%%%%%
\section{Gauge transformations of the spacetime}

A $U(1)$ gauge field $A_\mu$ with the constant field strength $F_{\mu\nu}$
can be written as
\bea
A_\mu = -\frac{1}{2} F_{\mu\nu}X^\nu\;.
\eea
Let $\det F_{\mu\nu} \neq 0$, therefore
\bea
X^\mu = -2 G^{\mu\nu} A_\nu\;, 
\eea
where $G = F^{-1}$. 
According to this equation we can perform gauge transformation on the 
spacetime coordinates. Under the gauge transformation
\bea
A_\mu \rightarrow {\tilde A}_\mu = A_\mu + \partial_\mu \Lambda \;,
\eea
the coordinate $X^\mu$ has the following transformation
\bea
X^\mu \rightarrow {\tilde X}^\mu = X^\mu - 2 G^{\mu\nu} 
\partial_\nu \Lambda\;. 
\eea
Therefore the gauge field ${\tilde A}_\mu$ is
\bea
{\tilde A}_\mu = -\frac{1}{2}F_{\mu\nu}{\tilde X}^\nu\;.
\eea
The equation (4) implies a transformation $X^\mu \rightarrow {\tilde X}^\mu 
=M^\mu_{\;\;\;\nu}X^\nu + a^\mu$ with the condition $FM+M^T F=2F$, can be 
interpreted as a gauge transformation in the 
spacetime with the gauge function 
\bea
\Lambda = \frac{1}{4}\sigma_{\mu\nu}X^\mu X^\nu
-\frac{1}{2}F_{\mu\nu}X^\mu a^\nu + \Lambda_0 \;,
\eea
where $\sigma=F({\bf 1}-M)$ is a symmetric matrix, ``$a^\mu$'' is a
constant vector and ``$\Lambda_0$'' is a constant number. 
Any translation $\delta X^\mu = a^\mu$ (i.e. $M={\bf 1}$) corresponds 
to a gauge transformation in the spacetime.
If the matrix $M$ is antisymmetric, the gauge function (6) describes
Lorentz transformations.

If the gauge function $\Lambda$ satisfies the equation
\bea
2(G^2)^{\mu\nu}\partial_\mu \Lambda \partial_\nu \Lambda + 
G^{\mu\nu}J_{\mu\nu} \Lambda = 0\;,
\eea
where $J_{\mu\nu}$ is the operator
$J_{\mu\nu} = X_\mu \partial_\nu - X_\nu \partial_\mu$,
the square length  $X^\mu X_\mu$ remains invariant. If $a^\mu =0$ and the
matrix $M$ is orthogonal, as expected the gauge function (6) satisfies 
this equation.

For a special gauge function $\Lambda$, it is possible to write 
${\tilde X}^\mu$ as a linear combination of $\{X^\mu\}$. Let the  
matrix $S_{\mu\nu}$ be symmetric and constant. For the gauge function 
\bea
\Lambda = \frac{1}{4}S_{\mu\nu}X^\mu X^\nu + \Lambda_0\;,
\eea
the gauge transformed coordinates are
\bea
{\tilde X}^\mu = ({\bf 1}-GS)^\mu_{\;\;\;\nu}X^\nu\;.
\eea

Now consider a rotation in the spacetime
\bea
X_{(r)}^\mu = {\cal{A}}^\mu_{\;\;\;\nu} X^\nu\;.
\eea
We do not restrict the matrix ${\cal{A}}$ to be orthogonal.
The gauge field $A_\mu$ transforms as
\bea
A_{(r)}^\mu = -\frac{1}{2}[({\cal{A}}^{-1})^T F]_{\mu\nu} X^\nu\;.
\eea
Equality of the rotated and gauge transformed versions
of the field $A_\mu$ leads to the equation
\bea
{\cal{A}}^{-1} = {\bf 1} + GS\;.
\eea
In addition if we impose the equality of the
gauged space and the rotated one, we obtain
\bea
{\cal{A}} = {\bf 1} - GS\;.
\eea
Therefore the matrices $S$ and $G$ should satisfy the condition
\bea
SGS = 0\;.
\eea
Thus we choose those gauge transformations with $\det S =0$, i.e. the
matrix $S$ is not invertible. For example 
for two dimensional spacetime the matrix $S$ has the form
\bea
S= \left( \begin{array}{cc}
a & \pm \sqrt{ab}\\
\pm \sqrt{ab} & b
\end{array} \right)\;.
\eea
For arbitrary $a$ and $b$ with $ab \geq 0$ this matrix satisfies the 
condition (14).

{\it Superstring in the gauged spacetime}

Now let the set $\{X^\mu\}$ denote the string coordinates.
The action of string that propagates in the background fields
$g_{\mu\nu}$, $A_\mu$ and the dilaton $\phi$, is \cite{5}
\bea
S&=&-\frac{1}{4\pi \alpha'} \int d^2 \sigma \sqrt{-h} \bigg{(}
h^{ab} g_{\mu\nu} \partial_a X^\mu\partial_b X^\nu
-\frac{\epsilon^{ab}}{\sqrt{-h}} F_{\mu\nu} 
\partial_a X^\mu\partial_b X^\nu 
\nonumber\\
&~&+ \alpha' R \phi(X) \bigg{)}\;.
\eea
Consider the gauge transformations of $g_{\mu\nu}$ and $\phi$ as
\bea
&~&g_{\mu\nu} \rightarrow {\tilde g}_{\mu\nu}\;,
\nonumber\\
&~&\phi \rightarrow {\tilde \phi}\;.
\eea
Invariance of the action (16) under the gauge transformations (9) and 
(17), leads to the equations
\bea
&~&{\cal{A}}^T {\tilde g}{\cal{A}} = g\;,
\nonumber\\
&~&{\cal{A}}^T F {\cal{A}} = F\;,
\nonumber\\
&~&{\tilde \phi} = \phi\;.
\eea
The set of matrices $\{{\cal{A}}\}$ that satisfy the equation
${\cal{A}}^T F {\cal{A}} = F$, form a group. According to the equation (13),
the second equation also gives the condition (14). 

The metric of the gauged space is
\bea
{\tilde g}_{\mu\nu} = g_{\mu\nu} +(GS-SG-SG^2S)_{\mu\nu}\;.
\eea
The second part purely is the effect of the chosen gauge.

Worldsheet supersymmetry gives the gauged fermions as
\bea
{\tilde \psi}_{\pm}^\mu(\tau \pm \sigma) = ({\bf 1}-GS)^\mu_{\;\;\;\nu}
\psi_{\pm}^\nu(\tau \pm \sigma)\;.
\eea
These are components of the fermions on the superstring worldsheet, that 
lives in the gauged spacetime. The first equation of (18)
implies the fermionic term of the superstring action, i.e.
$g_{\mu\nu} {\bar \psi}^\mu \rho^a \partial_a \psi^\nu$ is gauge invariant.
Therefore the superstring action is symmetric under the gauge transformations 
(9) and (18).

If some of the spacetime coordinates $\{X^{\bar \mu}\}$ are compact
on a tori with the radii $\{R_{\bar \mu}\}$, we have
\bea
&~&X^{\bar \mu}(\sigma+\pi , \tau)-X^{\bar \mu}(\sigma , \tau)
=2\pi L^{\bar \mu} = 2\pi n^{\bar \mu} R_{\bar \mu}\;,
\nonumber\\
&~&X'^{\bar \mu}(\sigma+\pi , \tau)-X'^{\bar \mu}(\sigma , \tau)
=2\pi \alpha' p^{\bar \mu} 
= 2 \pi m^{\bar \mu} \frac{\alpha'}{R_{\bar \mu}}\;,
\eea
where $X'^{\bar \mu}=X^{\bar \mu}_L-X^{\bar \mu}_R$ is 
$T$-dual coordinate of $X^{\bar \mu}$.
The sets $\{n^{\bar \mu}\}$ and $\{m^{\bar \mu}\}$ are winding and 
momentum numbers of the closed string around the compact directions 
$\{X^{\bar \mu}\}$. For the gauged coordinates we obtain
\bea
&~&{\tilde X}^{\bar \mu}(\sigma+\pi , \tau)-{\tilde X}^{\bar \mu}
(\sigma , \tau)= 2\pi {\tilde L}^{\bar \mu}\;,
\nonumber\\
&~&{\tilde X}'^{\bar \mu}(\sigma+\pi , \tau)-{\tilde X}'^{\bar \mu}
(\sigma , \tau)= 2\pi \alpha' {\tilde p}^{\bar \mu}\;,
\eea
where
\bea
&~&{\tilde L}^{\bar \mu} = ({\bf 1}-GS)^{\bar \mu}_{\;\;\;{\bar \nu}}
L^{\bar \nu}\;,
\nonumber\\
&~&{\tilde p}^{\bar \mu} = ({\bf 1}-GS)^{\bar \mu}_{\;\;\;{\bar \nu}}
p^{\bar \nu}\;.
\eea
If ${\tilde L}^{\bar \mu}$ is not zero, the coordinate ${\tilde X}^{\bar \mu}$ 
of the gauged space also is compact. Consider the case that the gauged
space is non-compact, i.e. ${\tilde L}^{\bar \mu}=0$ for any ${\bar \mu}$.
This leads to the condition $\det({\bf 1}-GS)=0$, which does not hold.
Therefore if some directions of the spacetime are compact, always some 
coordinates of the gauged space also are compact.
%%%%%%%%%%%%%%%%%%%%%%%%%%%%%%%%%%%%%%%%%%%%%%%%%%%%%%%%%%%%%%%%%%%%%%%%%%%%%
\section{General $T$-duality and gauge theory}

Consider the following transformations on the left and right 
moving parts of the compact coordinates $\{X^{\bar \mu}\}$ 
\bea
&~&X^{\bar \mu}_L(\tau+\sigma) \rightarrow Y^{\bar \mu}_L(\tau+\sigma)
={\cal{B}}^{\bar \mu}_{\;\;\;{\bar \nu}}X^{\bar \nu}_L(\tau+\sigma)\;,
\nonumber\\
&~&X^{\bar \mu}_R(\tau-\sigma) \rightarrow Y^{\bar \mu}_R(\tau-\sigma)
=-{\cal{B}}^{\bar \mu}_{\;\;\;{\bar \nu}}X^{\bar \nu}_R(\tau-\sigma)\;.
\eea
From these transformations we obtain
\bea
&~&X^{\bar \mu} \rightarrow Y^{\bar \mu}
={\cal{B}}^{\bar \mu}_{\;\;\;{\bar \nu}}X'^{\bar \nu}\;,
\nonumber\\
&~&X'^{\bar \mu} \rightarrow Y'^{\bar \mu}
={\cal{B}}^{\bar \mu}_{\;\;\;{\bar \nu}}X^{\bar \nu}\;.
\eea
Therefore the compact part of the spacetime and its $T$-dual, 
transform to each other.
For ${\cal{B}}={\bf 1}$ these are usual $T$-duality transformations.

Worldsheet supersymmetry enables us to obtain transformations of the  
worldsheet fermions, 
\bea
&~&\psi^{\bar \mu}_+ \rightarrow \chi^{\bar \mu}_+ =
{\cal{B}}^{\bar \mu}_{\;\;\;{\bar \nu}}\psi^{\bar \nu}_+ \;,
\nonumber\\
&~&\psi^{\bar \mu}_- \rightarrow \chi^{\bar \mu}_- =
-{\cal{B}}^{\bar \mu}_{\;\;\;{\bar \nu}}\psi^{\bar \nu}_- \;.
\eea
In other words we have the exchange transformations
\bea
&~&\psi^{\bar \mu} \rightarrow \chi^{\bar \mu}
={\cal{B}}^{\bar \mu}_{\;\;\;{\bar \nu}}\psi'^{\bar \nu}\;,
\nonumber\\
&~&\psi'^{\bar \mu} \rightarrow \chi'^{\bar \mu}
={\cal{B}}^{\bar \mu}_{\;\;\;{\bar \nu}}\psi^{\bar \nu}\;.
\eea
That is, the worldsheet fermions $\{\psi^{\bar \mu}\}$ and their
$T$-dual fermions $\{\psi'^{\bar \mu}\}$ are exchanged.

The invariance of the mass spectrum of the closed superstring leads to the
fact that the matrix ${\cal{B}}$ is orthogonal
\bea
{\cal{B}}^T {\cal{B}} = {\cal{B}}{\cal{B}}^T = {\bf 1}\;.
\eea

Compactification of the $\{X^{{\bar \mu}}\}$ directions leads to the
compactification of the coordinates $\{Y^{\bar \mu}\}$.
Using the equations (21), gives
\bea
&~& Y^{\bar \mu}(\sigma+\pi , \tau)-Y^{\bar \mu}(\sigma , \tau)
=2\pi \Lambda^{\bar \mu}\;,
\nonumber\\
&~&Y'^{\bar \mu}(\sigma+\pi , \tau)-Y'^{\bar \mu}(\sigma , \tau)
=2\pi \alpha' \Pi^{\bar \mu}\;,
\eea
where $\Pi^{\bar \mu}$ and $\Lambda^{\bar \mu}$ are
\bea
&~&\Pi^{{\bar \mu}} = \frac{1}{\alpha'} {\cal{B}}^{{\bar \mu}}
_{\;\;\;{\bar \nu}}L^{{\bar \nu}}\;,
\nonumber\\
&~& \frac{1}{\alpha'}\Lambda^{{\bar \mu}} = {\cal{B}}^{{\bar \mu}}
_{\;\;\;{\bar \nu}} p^{{\bar \nu}}\;.
\eea
These equations imply the momentum and the winding numbers of closed string
in the Y-space depend on the winding and momentum 
numbers of it in the compact part of the spacetime respectively.

Previously we have observed that for an emitted closed string  
from a D$_p$-brane with the background fields $A_\mu$ and $B_{\mu\nu}$
(the NS$\otimes$NS field),
there is the following relation between its momentum and winding numbers
\cite{4}
\bea
p^\alpha = -\frac{1}{\alpha'}{\cal{F}}^\alpha_{\;\;\;\beta}L^\beta\;,
\eea
where ${\cal{F}}_{\mu\nu}=F_{\mu\nu}-B_{\mu\nu}$ is
total field strength and the indices $\alpha$ and 
$\beta$ show the brane directions. According to this relation we have
\bea
\frac{1}{\alpha'}\Lambda^{\bar \alpha}=-({\cal{B}}{\cal{F}}{\cal{B}}^T)
^{\bar \alpha}_{\;\;\;{\bar \beta}}\Pi^{\bar \beta}\;.
\eea
Comparing this equation with the relation (31) implies, in the Y-space
closed string feels the total field strength ${\cal{B}}{\cal{F}}^{-1}
{\cal{B}}^T$.

{\it Transformation of the string action}

The action of a string that propagates in the background fields 
$g_{\mu\nu}$, ${\cal{F}}_{\mu\nu}$ and $\phi$, is \cite{5}
\bea
S&=&-\frac{1}{4\pi \alpha'} \int d^2 \sigma \sqrt{-h} \bigg{(}
h^{ab} g_{\mu\nu} \partial_a X^\mu\partial_b X^\nu
-\frac{\epsilon^{ab}}{\sqrt{-h}}{\cal{F}}_{\mu\nu} 
\partial_a X^\mu\partial_b X^\nu 
\nonumber\\
&~&+ \alpha' R \phi(X) \bigg{)}\;.
\eea
In addition to the general $T$-duality transformations (25), consider 
the following transformations for the background fields 
\bea
&~& g_{\mu\nu}\rightarrow {\bar g}_{\mu\nu} \;,
\nonumber\\
&~& {\cal{F}}_{\mu\nu}\rightarrow {\bar {\cal{F}}}_{\mu\nu} \;,
\nonumber\\
&~& \phi \rightarrow {\bar \phi}\;.
\eea
Assume that all directions of the spacetime are compact on tori.
Therefore the action (33) transforms as
\bea
{\bar S}&=&-\frac{1}{4\pi \alpha'} \int d^2 \sigma \sqrt{-h} \bigg{(}
h^{ab} ({\cal{B}}^T{\bar g}{\cal{B}})_{\mu\nu} \partial_a 
X'^\mu\partial_b X'^\nu
-\frac{\epsilon^{ab}}{\sqrt{-h}} ({\cal{B}}^T{\bar {\cal{F}}}
{\cal{B}}) _{\mu\nu} 
\partial_a X'^\mu\partial_b X'^\nu 
\nonumber\\
&~&+ \alpha' R {\bar \phi}(X') \bigg{)}\;.
\eea

Invariance of the action under the transformations (25) and (34) 
leads to the relations
\bea
&~&{\cal{B}}^T {\bar g}{\cal{B}} = g'\;,
\nonumber\\
&~&{\cal{B}}^T {\bar {\cal{F}}}{\cal{B}} = {\cal{F}}'\;,
\nonumber\\
&~&{\bar \phi} = \phi'\;,
\eea
where $g'_{\mu\nu}$, ${\cal{F'}}_{\mu\nu}$ and $\phi'$ denote the 
usual $T$-duality of these background fields. 
These equations give ${\bar g}={\cal{B}}g'{\cal{B}}^T $ and
${\bar {\cal{F}}}={\cal{B}} {\cal{F'}} {\cal{B}}^T $.
Therefore ${\bar g}_{\mu\nu}$, 
${\bar {\cal{F}}}_{\mu\nu}$ and ${\bar \phi}$ are general $T$-duality 
transformations of the background fields, that are consistent with the
equations (25). In fact the equations (36) change the action (35) to 
the $T$-dual form of the action (33). 

We know that under the gauge transformations the total field strength
${\cal{F}}$ is invariant. We are interested in the case that above 
general $T$-duality to be a gauge transformation. 
In this case ${\bar {\cal{F}}}$ is proportional to ${\cal{F}}$, i.e.
\bea
{\bar {\cal{F}}}_{\mu\nu} = -\frac{1}{\lambda}{\cal{F}}_{\mu\nu}\;,
\eea
where $\lambda$ is a constant number. The $T$-duality 
of the total field strength ${\cal{F}}$ is \cite{6}
\bea
{\cal{F}}' = -(g+k{\cal{F}})^{-1}{\cal{F}}(g-k{\cal{F}})^{-1}\;,
\eea
where $k=\pm1$. According to the equations (36)-(38) we can obtain the 
matrix ${\cal{B}}$,
\bea
{\cal{B}} = \ell\sqrt{\lambda} \Gamma (g-k{\cal{F}})^{-1}\;,
\eea
where $\ell=\pm 1$. The matrix $\Gamma$ is limited as the following
\bea
\Gamma^T {\cal{F}} \Gamma = {\cal{F}}\;.
\eea
The orthogonality of the matrix ${\cal{B}}$ leads to the orthogonality 
of the matrix $\Gamma$
\bea
\Gamma^T \Gamma = \Gamma \Gamma^T = {\bf 1}\;,
\eea
and an additional condition on the background fields 
\bea
g^2 - {\cal{F}}^2 = \lambda {\bf 1}\;.
\eea
This condition means that  
only some special background fields admit the equation (37) to hold.
The set of orthogonal $\Gamma$ matrices that satisfy the equation (40) form 
group $O(d)$, where ``$d$'' is the spacetime dimension.

Since $T$-duality of the metric $g_{\mu\nu}$ is 
$g' = (g+k{\cal{F}})^{-1}g(g-k{\cal{F}})^{-1}$,
the equations (39) and (42) give the metric ${\bar g}$ as
\bea
{\bar g} = \frac{1}{\lambda} \Gamma g \Gamma^T\;.
\eea
This is gauge transformed version of the spacetime metric.
For $\Gamma = {\bf 1}$ and $\lambda = -1$ the theory is gauge invariant 
and ${\bar g}= -g$. That is the signature of the metric is changed.

According to the equation (39), the relation (32) can be written as
\bea
\frac{1}{\alpha'} \Lambda ^\alpha = \lambda (\Gamma {\cal{F}}'
\Gamma^T)^{\alpha}_{\;\;\; \beta} \Pi^\beta\;.
\eea
That is, in the gauged space closed string feels the effective field
strength $-\frac{1}{\lambda} \Gamma {\cal{F'}}^{-1} \Gamma^T$.
%%%%%%%%%%%%%%%%%%%%%%%%%%%%%%%%%%%%%%%%%%%%%%%%%%%%%%%%%%%%%%%%%%%%%%%%%%%%%
\section{Conclusion}
We observed that in the presence of a $U(1)$ gauge field 
with constant field strength, it is possible
to perform gauge transformations on the spacetime coordinates, the metric
$g_{\mu\nu}$ and the dilaton $\phi$. There are
some gauge transformations that preserve the spacetime distances. 
Some rotations and translations in the spacetime are equivalent to some
gauge transformations. For example Lorentz transformation is one of them.
We showed that the string action under a special gauge 
transformation is invariant. In addition we saw compactification of the
spacetime is induced to the gauged space.

We showed that there is a generalization of the 
$T$-duality that preserves the
compactification of the spacetime. In this duality the winding and momentum
numbers of closed string get exchange. Invariance of the string action
with the NS$\otimes$NS background fields under the general $T$-duality
gives general $T$-duality
transformations of these fields. For special background fields the
above general $T$-duality transformations are equivalent to gauge 
transformations. In this gauged space closed string probes a modified field
strength.
%%%%%%%%%%%%%%%%%%%%%%%%%%%%%%%%%%%%%%%%%%%%%%%%%%%%%%%%%%%%%%%%%%%%%%%%%%%%


\begin{thebibliography}{99}
\bibitem{1}
A.M. Polyakov, ``{\it Gauge Fields and Space-Time}'', hep-th/0110196.
\bibitem{2}
M. Dine, P. Huet and N. Seiberg, Nucl. Phys.{\bf B322}(1989)301.
\bibitem{3}
D. Kamani, Nucl. Phys. {\bf B601}(2001)149-168, hep-th/0104089.
\bibitem{4}
H. Arfaei and D. Kamani, Phys. Lett.{\bf B452}(1999)54, hep-th/9909167. 
\bibitem{5}
R.G. Leigh, Mod. Phys. Lett.{\bf A4} 28(1989)2767.
\bibitem{6}
A. Giveon, M. Porrati, E. Rabinovici, Phys. Rep. 244(1994)77-202.

\end{thebibliography}
\end{document}